\documentclass[format=sigconf, nonacm]{acmart}
\AtBeginDocument{%
  }

\usepackage{graphicx} 
\graphicspath{{figures/}}
\usepackage[ruled, linesnumbered]{algorithm2e}
\usepackage{algpseudocode}

\begin{document}

\title{Utility-aware Social Network Anonymization \\ using Genetic Algorithms}

\author{Samuel Bonello, Rachel G. de Jong, Thomas H. W. B\"ack and Frank W. Takes}
\email{r.g.de.jong@liacs.leidenuniv.nl}
\affiliation{%
  \institution{Universiteit Leiden}
  \country{The Netherlands}
}

\renewcommand{\shortauthors}{S. Bonello, R. G. de Jong, T. H. W. B\"ack and F. W. Takes}

\begin{abstract}
    Social networks may contain privacy-sensitive information about individuals. 
The objective of the network anonymization problem is to alter a given social network dataset such that the number of anonymous nodes in the social graph is maximized. 
Here, a node is anonymous if it does not have a unique surrounding network structure. 
At the same time, the aim is to ensure data utility, i.e., preserve topological network properties and retain good performance on downstream network analysis tasks. 
We propose two versions of a genetic algorithm tailored to this problem: one generic GA and a uniqueness-aware GA (UGA). 
The latter aims to target edges more effectively during mutation by avoiding edges connected to already anonymous nodes. 
After hyperparameter tuning, we compare the two GAs against two existing baseline algorithms on several real-world network datasets. 
Results show that the proposed genetic algorithms manage to anonymize on average 14 times more nodes than the best baseline algorithm. 
Additionally, data utility experiments demonstrate how the UGA requires fewer edge deletions, and how our GAs and the baselines retain performance on downstream tasks equally well. 
Overall, our results suggest that genetic algorithms are a promising approach for finding solutions to the network anonymization problem. 
\end{abstract}



\maketitle

\section{Introduction}\label{sec:intro}
In the field of network science or social network analysis, real-world social network data is frequently used, for example to model networked processes such as influence maximization~\cite{li2018influence} or epidemic spread~\cite{azizi2020epidemics}. 
Here, the social network data is a graph in which the nodes are people, and edges represent meaningful connections such as friendship or human communication. 
Network data is typically the input of downstream network analysis tasks and algorithms, such as assessing network robustness~\cite{albert2000error}, finding central nodes~\cite{saxena2020centrality} and community detection~\cite{blondel2008fast}.
In light of open science, social network data would ideally always be openly accessible. 
However, in many situations, a social network contains personally identifiable information that upon disclosure would violate people's privacy.

Various more generic approaches for minimizing above mentioned disclosure risk have been introduced in literature, for which the focus was initially on more traditional tabular data~\cite{hundepool2012statistical, sweeney2002k, dwork_differential_2008}.
However, network data introduces new challenges and requires different approaches as the network structure surrounding a node can itself be used to identify individuals~\cite{backstrom2007wherefore, romanini2021privacy, dejong2023effect}.
Existing research to mitigate this problem can roughly be categorized into three streams of work, concretely being differential privacy~\cite{sala2011sharing, proserpio2012calibrating, wang2013preserving}, clustering methods~\cite{campan2008data, bhagat2009class, liu2016linkmirage, minello2020k} and approaches based on $k$-anonymity~\cite{liu2008towards, hay2008resisting, romanini2021privacy, zou2009k}. We briefly discuss all three below. 

In differential privacy, the user does not get access to the complete dataset, but instead is allowed to query parts of the data. 
Privacy is then guaranteed by adding noise to query answers in such a way that the presence or absence of an individual can not be derived.
Some network-specific differential privacy approaches focus on answering a single query~\cite{proserpio2012calibrating, hay2009boosting}, while others generate a graph based on the answer produced, such as the joint degree distribution~\cite{sala2011sharing, wang2013preserving, xiao2014differentially}.
In the second category of approaches based on clustering, privacy is ensured by creating a new network in which nodes are grouped together into large enough supernodes that no longer reveal the identity of the nodes contained in it. 

A drawback of these two categories of approaches is that either a user is limited by which queries can be asked, or certain graph properties are substantially distorted.
For example, in differential privacy communities are not guaranteed to be preserved, as the network is generated from a global distorted joint degree distribution. 
More generally, these approaches often lead to a loss or even impossibility of measuring \emph{data utility}, i.e., the extent to which topological properties of the network are preserved, and performance on downstream tasks is retained. 
Additionally, in many fields of applied research, such as the social sciences, it must be possible to publish the network dataset underlying an analysis for reasons of reproducibility. 
Therefore, in this paper we focus on the third category, being $k$-anonymity, as this approach ultimately enables publication of (parts of) a $k$-anonymized version of the network.

A node is said to be $k$-anonymous if it is equivalent to at least $k-1$ other nodes.
In this paper, we focus on the scenario in which a node is anonymous if it is not unique, i.e., $k=2$.
The criteria for two nodes to be equivalent, depends on the chosen anonymity measure.  
This measure essentially represents the attacker scenario, i.e., what information is in the hands of a potential adversary aiming to deanonymize the network. 
In the version of the \emph{network anonymization problem} considered in this paper, the aim is to maximize the number of 2-anonymous nodes with a limited number of alterations to the graph. 
Based on the chosen anonymity measure the difficulty of anonymizing a network in terms of computational complexity differs.
For example, when using the degree of a node as anonymity measure, the problem can be solved in $\mathcal{O}(n^2)$ time, where $n$ is the number of nodes~\cite{liu2008towards}. 
However, when the measure takes into account the exact graph structure surrounding a node, this problem is NP-hard~\cite{zhou2008preserving}, requiring approximation algorithms to handle larger graph inputs.

The anonymization problem is not monotonic~\cite{dejong2024anonymization}, implying that more alterations do not always result in higher anonymity.
Moreover, the anonymization problem is clearly an example of an optimization problem with a tremendously large search space, as a subset of all edges should be selected for modification in order to maximize the number of anonymous nodes.
Hence, a genetic algorithm is a logical candidate for the anonymity maximization problem.
Moreover, GAs have proven to be effective in other problems where the input is a (social) network~\cite{lotf2022improved, azaouzi2019community, caschera2019monde}.

\begin{figure*}[!t]
	\centering
    \vspace{-1em}
    \includegraphics[width=\textwidth]{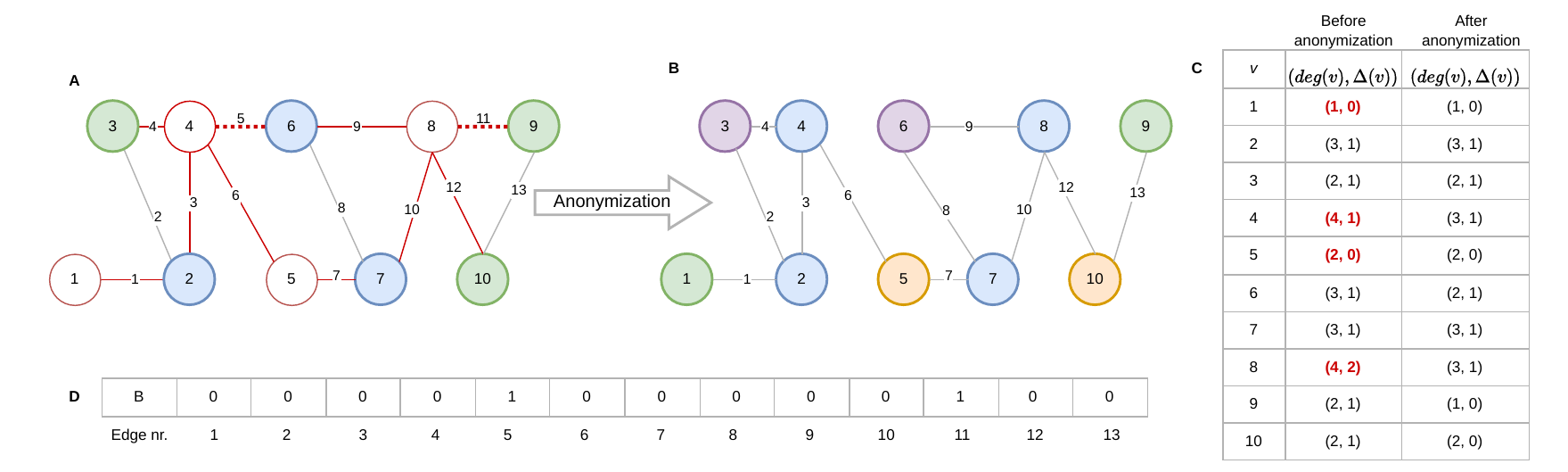}
    \vspace{-1em}
	\caption{Example of the network anonymization process. Figure A: Toy graph where nodes with the same color are equivalent, i.e., have the same degree and are part of the same number of triangles, and red nodes are unique (not anonymous). Red edges are unique (connect to a unique node), and dotted edges are deleted in the anonymization process. Figure B: Anonymized graph where each node is anonymous. Figure C: State of each node before and after anonymization according to the \textsc{count} measure, with unique states shown in red. Figure D: The bitstring of modified edges encoding this solution.}
	\label{fig:anonymizing}
\end{figure*}

In this paper, we first introduce a generic genetic algorithm (GA) tailed to the network anonymization problem. 
An individual in the population is a bitstring of length $m$, where $m$ is the number of edges in the network, and mutation from $0$ to $1$ indicates that in the solution, this edge is deleted from the graph. 
Here, we focus on deletion, which compared to addition or rewiring, was shown in previous work to best improve anonymity~\cite{dejong2024anonymization} as this operation makes neighborhoods of nodes smaller, resulting in less variety in neighborhood structures and thus on average more anonymity.
Additionally, allowing for edge addition in a genetic algorithm would increase the size of the individual to all possible edges, $n \choose 2$, where $n$ equals the number of nodes in the undirected graph. 
Figure~\ref{fig:anonymizing} shows an example of anonymization by means of edge deletion. 
Second, we propose a uniqueness-aware genetic algorithm (UGA) that in its mutation step explicitly avoids deleting edges that connect already anonymous nodes. 
Additionally, we account for data utility by implementing a constraint on the maximum number of edges deleted.
Our focus is on proposing anonymity measure-agnostic GAs. 
As opposed to previously proposed evolutionary approaches that were either tailored to a specific type of network analysis problem or anonymity measure \cite{alavi2019attacker, rajabzadeh2020graph}, our algorithms can be used for a range of anonymity measures and (social) network datasets.
Therewith they provide flexibility in terms of modeling different attacker scenarios according to the user's preferences. 

To summarize, the contributions of this paper are as follows.
\begin{itemize}
    \item We propose novel scalable and efficient heuristic network anonymization algorithms in the form of a generic and a uniqueness-aware genetic algorithm (GA and UGA). 
    \item We compare the performance of the two newly proposed GAs to two existing baseline algorithms from the literature, using real-world social network datasets, showing how proposed GAs perform ca. 14 times better than the best baseline. 
    \item We critically assess the performance/utility trade-off of the proposed GAs, and demonstrate that they preserve or even improve data utility and downstream task performance. 
\end{itemize}
The remainder of this paper is structured as follows.
First, Section~\ref{sec:rel} gives an overview of related work on anonymization of networks.
Next, Section~\ref{sec:prel} summarizes concepts and notation used in the remainder of the paper.
In Section~\ref{sec:approach}, we introduce the genetic algorithms, which we compare against state of the art in Section~\ref{sec:experiments}.
Lastly, we give a summary and main conclusions in Section~\ref{sec:conc}.

\section{Related work}\label{sec:rel}
Various approaches for the anonymization of networks have been introduced in the literature.
An important choice in each of these approaches, as mentioned in Section~\ref{sec:intro},  is the choice of the anonymity measure, that is, the requirement for when two nodes are considered to be equivalent. 
Interestingly, one of the findings in recently published comparative work~\cite{dejong2024comparison} is that a relatively simple and easy to compute node anonymity measure based on the number of directly surrounding nodes and triangles, in real-world scenarios often provides results similar to those of measures that use the exact structure of the node's direct neighborhood. 
Hence, we choose to use this anonymity measure, as further detailed in Section~\ref{sec:prel}. 

Many previously introduced works are what we refer to as measure-specific approaches. Below, we discuss them in more detail, distinguishing between exact and approximate algorithms.
The work in~\cite{liu2008towards} introduces an algorithm that anonymizes a given network in the smallest number of edge alterations in quadratic time, in terms of the number of nodes. 
However, this is a specific result for the case where the anonymity measure used is the degree of the nodes, and both edge deletion and addition are allowed.
In a different scenario, when using a more complex anonymity measure and only edge deletion, the problem is NP-hard~\cite{zhou2008preserving}. Hence, an exact approach would not be suitable for larger graphs.

Greedy approaches have also been introduced. 
The work in~\cite{zhou2008preserving} assumes that nodes are equivalent when their neighborhoods are isomorphic. The corresponding anonymization algorithm uses an approximation method to make the neighborhoods of targeted nodes isomorphic. 
Following an even more strict measure, anonymity is achieved if the graph can be partitioned into $k$ isomorphic parts~\cite{zou2009k}. In this case, the graph is anonymized by partitioning it into $k$ segments and altering the graph such that these segments are isomorphic.
Both of these approaches are measure-specific and may not use the minimum number of edge alterations.

Measure-agnostic approximation algorithms have also been introduced in recent literature~\cite{dejong2024comparison}. 
This body of work consists of approximation approaches for anonymization, for example, using random edge sampling~\cite{romanini2021privacy}, or swapping~\cite{hay2007anonymizing}.
Notable, various heuristic based approaches aiming to target edges with particular topological network properties have been introduced in~\cite{dejong2024anonymization}.

The anonymization problem in networks is an example of an optimization problem operating on graph data. 
It is well-known that GAs are effective in solving other optimization problems in networks, such as influence maximization~\cite{lotf2022improved}, community detection~\cite{azaouzi2019community} or predicting dynamics of a social network~\cite{caschera2019monde}. 
For network anonymization, only a few studies have considered the use of GAs. 
However, both are tailored to a specific anonymity measure and either use a range of graph modification operations~\cite{alavi2019attacker}, or aim to retain performance on a specific network analysis task such as community detection~\cite{rajabzadeh2020graph}.

Instead, in this work we focus on proposing measure-agnostic genetic algorithms, with the aim of scaling to networks with thousands of nodes, overcoming the limitations of exact approaches, while retaining control over data utility and ensuring performance on downstream network analysis tasks. 
Hence, our baselines for comparison are edge sampling~\cite{romanini2021privacy} and the most effective heuristic algorithm presented in~\cite{dejong2024anonymization}. 

\section{Preliminaries}\label{sec:prel}
In this section, we summarize the concepts and notation on networks and anonymity used throughout the remainder of the paper.

\subsection{Networks}
\label{sec:networks}
A network $G = (V, E)$ consists of a set of nodes $V$ and a set of edges $\{v, w\} \in E$ connecting pairs of nodes $v,w \in V$. 
While model extensions to include directionality, weights and attributes exist, here we focus on networks modeled as undirected unweighted graphs and hence, an edge is a set of size two, and not a tuple. 
The degree of a node, $deg(v)$ equals the number of connections of node $v \in V$.
For a node $v$, the set of neighbors $N(v)$ equals the set of directly connected nodes, i.e., $N(v) = 
\{w : \{v, w\} \in E\}$. Thus, $deg(v) = |N(v)|$. 
Neighboring nodes often form triangles.
The number of triangles a node is part of is $\Delta(v) = |\{ \{u, w\} \in E : u, w \in N(v)\}|$.
The extent to which neighboring nodes form triangles is captured by the node clustering coefficient $c(v)$, being 

\begin{align}  
c(v) = \frac{\Delta(v)}{\frac{1}{2} deg(v) * (deg(v) - 1) }
\label{eq:clust}
\end{align}
The graph's average clustering coefficient $cc(G)$ is then equal to $\frac{1}{|V|} \sum_{v \in V} c(v)$, and describes the overall tendency of nodes to cluster together. 
The distance between two nodes, denoted $d(v, w)$, equals the minimum number of edges that must be traversed to reach node $w$, starting from node $v$.   
As the graph is undirected, $d(v, w) = d(w, v)$.
The diameter is the maximal finite distance over all node pairs. 
If there is no path between two nodes, we define $d(v, w) = \infty$.
When computing the average distance over all node pairs in the graph, these values are ignored.
This occurs when the nodes are in different \emph{components}, i.e., a set of nodes such that for all node pairs $v$ and $w$ in this component,  $d(v, w) \neq \infty$.
In real-world networks, usually the vast majority of nodes are in the largest connected component (LCC) of the graph, also called giant component, which captures the vast majority of connectivity~\cite{barabasi2016network}. 
Moreover, real-world networks are usually relatively sparse and have low average distances, an observation in real-world network data known as the ``small-world phenomenon''~\cite{kleinberg2000small}.  

In real-world networks, nodes often form communities, being clusters of densely connected nodes.  A partitioning of the graph can be found using community detection algorithms. As even the most commonly used approaches suffer from instability due to randomization effects, consensus clustering can be used~\cite{lancichinetti2012consensus} to find a stable division of the network into communities.
Nodes can have different roles or importance in a network, with some nodes being more central than others. In this paper we capture the centrality of a node using betweenness centrality, one of the most common nontrivial centrality measures proposed in the literature~\cite{saxena2020centrality}. 
For a given node, it measures the fraction of shortest paths going through that node~\cite{brandes2001faster}.

\subsection{Measuring anonymity}
\label{sub:measuring}
While many different anonymity measures have been introduced in the literature~\cite{dejong2024comparison}, as motivated in Section~\ref{sec:rel}, in this work we focus on the local node measure referred to in previous work as \textsc{count}.
With this measure, two nodes are equivalent if they have the same degree, and are part of the same number of triangles. 
Together, this is referred to as \emph{state} of the node. An example is shown in Figure~\ref{fig:anonymizing}C.
Note that while various extensions to the network data model can be made, such as adding node or edge labels, timestamps or edge directionality, for which of course the anonymity measures should be extended accordingly, in this paper we choose to focus on the undirected network, accounting only for structure. 

If a node is equivalent to $k-1$ nodes, i.e., there are $k$ nodes in $V$ with this state, it is $k$-anonymous.
In this work, we focus on the case $k=2$, so a node is anonymous if it is equivalent to at least one other node, and hence not unique.
The set of unique nodes is denoted $V_u$.
The set of unique edges connecting at least one unique node is $E_u$. 
Both are defined below. 
\begin{align}
    V_u = \{v \in V : \nexists_{w \neq v \in V} : deg(v) = deg(w) \wedge \Delta(v) = \Delta(w)\}  \label{eq:vu}
\end{align}
\begin{align}
    E_u = \{ \{v, w\} \in E : v \in V_u \vee w \in V_u \} \label{eq:eu}
\end{align}
Following \cite{romanini2021privacy}, we maximize anonymity $1 - U(G)$ by minimizing uniqueness $U(G)$, i.e., the fraction of unique nodes in the network: 
\begin{align}
    U(G) = \frac{|V_u|}{|V|} \label{eq:uniq}
\end{align}

\subsection{Anonymization problem}\label{sub:anon}
The goal of anonymization is to minimize the uniqueness of the network by means of perturbation, in our case focusing on edge deletion, which, as also argued in Section~\ref{sec:intro}, was shown in previous work~\cite{dejong2024anonymization} to be a promising operation. More importantly, it limits the search space to the set of all existing edges. To retain data utility, we introduce a budget $\Gamma$, explicitly limiting the number of deletions.
Figure~\ref{fig:anonymizing} shows an example of a complete solution to the anonymization problem by means of edge deletion.
We formally define the problem as follows.

\begin{definition}[Anonymization problem.]
    Given a network $G$ and budget $\Gamma$, delete a set of edges $E_{del} \subset E$, with $|E_{del}| \leq \Gamma$, maximizing anonymity $1 - U(G)$, while minimizing $|E_{del}|$. 
\label{def:ano}
\end{definition}
To compute the number of unique nodes, we must determine the state of each node as defined in Section~\ref{sub:measuring}, i.e., the degree and number of triangles. 
 The complexity of this is determined by the latter, which can be done in $\mathcal{O}(|V|^2)$ for each node. 

\subsection{Data utility measures and downstream tasks}
\label{sub:utility}

Although the budget $\Gamma$ defined above limits the number of edges that can be deleted, this does not guarantee that derived topological network properties are preserved, nor that performance on downstream network analysis tasks is retained. 
Therefore, to measure the utility of the anonymized graph, we compare three topological network properties and three downstream tasks. Each of these properties and downstream tasks are commonly studied in the fields of social network analysis and network science~\cite{barabasi2016network}. The topological properties include: 
\begin{enumerate}
    
    \item The actual number of edges deleted: an obtained solution might require fewer than the allowed $\Gamma$ edge deletions for an equally optimal result, retaining more of the original social network data in the anonymized network. 
    
    \item The change in the graph's average clustering coefficient $cc(G)$ (defined in Section~\ref{sec:networks}). 
    As edges are deleted, triangles will be destroyed, while at the same time decreasing the degree of nodes.
    As a result, the average clustering coefficient is expected to decrease.
    
    \item The average shortest path length $d(v,w)$ over all node pairs $v, w \in V$ in the same connected component. 
    Given that we focus on relatively small budgets $\Gamma$, we expect a giant component to remain intact, meaning that distances would only increase as edges are deleted. 
\end{enumerate}
The considered downstream tasks are as follows: 
\begin{enumerate}
    \item Network robustness: Measured as the fraction of nodes in the largest connected component (LCC).
    As edges are deleted and nodes become disconnected from the giant component, this number is expected to decrease.
    \item Centrality: Measured as the similarity of the node sets forming the top 100 most central nodes, according to betweenness centrality (see Section~\ref{sec:networks}) before and after anonymization.
    This number equals 1.0 if the top 100 did not change, and otherwise decreases as central nodes miss in the top 100 of the anonymized graph.
    \item Community structure: Measured as the normalized mutual information (NMI)~\cite{lancichinetti2012consensus}
 between the communities found before and after anonymization. 
    NMI equals 1.0 if the communities are the same as before anonymization, and is expected to decrease as the network is perturbed during anonymization.
\end{enumerate}
\section{Approach}\label{sec:approach}
In this section, we introduce the proposed genetic algorithms. 
First, we describe the representation of individuals in the population and the objective function modeling the network anonymization problem. Then, we introduce both the generic genetic algorithm (GA) and uniqueness-aware GA (UGA) and their genetic operators. 

\subsection{Encoding of an individual}\label{sub:ind}


In network anonymization, a solution is a modified version of the input graph, where certain edges are deleted. 
To encode such a solution, each individual is a bitstring $B = (b_1,\ldots,b_{|E|})$ of length $|E|$. Assuming edges (and bits) are numbered from $1$ to $|E|$, the $i^{th}$ bit $b_i$ equals 1 if the edge will be deleted, and 0 if the edge is preserved.
This representation is also illustrated in Figure~\ref{fig:anonymizing}D.

\subsection{Optimization problem}
\label{sub:opt}

An obvious solution to make all nodes anonymous would be to delete all edges.
However, in this case all information of the original network, i.e., data utility, is lost.
Hence, to account for utility, we formulate an optimization problem based on on the network anonymization problem in Definition~\ref{def:ano} that includes a constraint to limit the number of deleted edges, as overall deleting fewer edges results in higher data utility. 
This results in the following optimization problem, where $G'_B$ is the graph $G$ from which the edges marked $1$ in bitstring $B$ (modeling $E_{del}$ in  Definition~\ref{def:ano}) are deleted: 
\begin{eqnarray}
    \min_{B \in \{0,1\}^{|E|}} U(G'_B) & & \\ \label{eq:opt}
        \text{subject to} \sum_i^{|E|} b_i \leq \Gamma & & \label{eq:const}
\end{eqnarray}

\subsection{Objective function}
To encode the optimization problem formulated above and adequately account for the limit on the number of edges that is allowed to be deleted, we introduce a penalty.
This increases the objective function value when the number of edges deleted is larger than the budget.
The penalty is equal to the number of edges by which it exceeds the budget $\Gamma$, and 0 if the limit is not exceeded.
Altogether, this results in the following penalty and objective function which we aim to minimize:
\begin{eqnarray}
    penalty & = & \max(0.0, \ \sum_i^{|E|} b_i - \Gamma) \\ \label{eq:penalty}
    f(G'_B) & = & |V_u| + penalty \; \rightarrow \; \min \label{eq:fitness} 
\end{eqnarray}

\subsection{Generic Genetic Algorithm (GA)}\label{sub:basic}

Below we summarize the workings of the proposed  genetic algorithm. The pseudocode can be found in Algorithm~\ref{alg:ga}, and an overview of hyperparameters used is shown in Table~\ref{tab:params}.

The first step in line~\ref{alg:ga:init} generates the initial population.
The function initializes the $\mu$ individuals by setting each bit to 1 with a probability $p$.
Next, line~\ref{alg:ga:fit1} determines the objective function values according to  Equation~\ref{eq:fitness}.
This operation includes determining the number of unique nodes. 
In line~\ref{alg:ga:best}, the best individual and objective function value in the initial population are determined.
The while-loop in lines~\ref{alg:ga:whliles} to~\ref{alg:ga:whilee} performs the main part of the genetic algorithm.
First, parental selection is performed by means of roulette wheel (i.e., proportional) selection.
For this step, to ensure that better solutions have a higher probability of being selected, the probability of selecting an individual $j$ equals
\begin{align}
    P(B_j) = ({\max}_f - f(G'_{B_j}))/\sum_{i=1}^\lambda ({\max}_f - f(G'_{B_i}))
\end{align}
where $B_i$ is the $i$-th individual in the population and $\max_f$ the largest objective function value of the current population.

In line~\ref{alg:ga:cross}, the chosen parents form new children by performing crossover.
This is either uniform, or with $c$ crossover points.
The new population undergoes a mutation step in which each bit of each individual is flipped with probability $\alpha$.
In lines~\ref{alg:ga:fit2} to~\ref{alg:ga:best2-end} the objective function values of the new population are computed and the best individual and value found so far are updated.
Thereafter, in line~\ref{alg:ga:sel}, the new population is selected by means of $(\mu + \lambda)$-selection.

As the last step of the algorithm's main loop, the mutation rate is updated according to Equation~\ref{eq:decay}.
By using mutation decay rate $\eta$, the mutation rate changes over time.
As a result, the search becomes more ``local'' as the number of generations increases.
The maximum of the two terms ensures that the mutation will, on average, flip at least one bit for an individual.
\begin{align}
    \alpha = \max(\alpha(1 - \eta \cdot gen), \frac{1}{|E|}) \label{eq:decay} 
\end{align}

The algorithm terminates when either 1) no better individual has been found for $\tau$ iterations or 2) an individual with objective function value 0 has been found.
After termination, line~\ref{alg:ga:return} returns the best individual found and its objective function value.

\begin{algorithm}[t]
    \textbf{Input: graph $G=(V, E)$, hyperparameters $\mu$, $p$, $\lambda$, $c$, $\alpha$, $\eta$, $S_p$, $S_e$, $\Gamma$, $\tau$}     

    $pop \leftarrow initialize(|E|,\ \mu,\ p)$ \\ \label{alg:ga:init}
    $f\_pop \leftarrow f(G,\ pop,\ \Gamma)$ \\ \label{alg:ga:fit1}
    $ind\_best,\ f\_best \leftarrow get\_best(pop,\ f\_pop)$ \\ \label{alg:ga:best} 
    $gen \leftarrow 0$, \ $gen\_best \leftarrow 0$ \\ \label{alg:ga:geninit}
    \While{$gen - gen\_best < \tau$ \textbf{and} $f\_best \neq 0$}{ \label{alg:ga:whliles}
        $parents \leftarrow parental\_selection(pop,\ f\_pop,\ \lambda,\ S_p)$ \\ \label{alg:ga:ps}
        $pop' \leftarrow crossover(parents,\ \lambda,\ c)$ \\ \label{alg:ga:cross}
        $pop' \leftarrow mutate(pop',\ \alpha)$ \\ \label{alg:ga:mut}
        $f\_pop' \leftarrow f(G,\ pop',\ \Gamma)$ \\ \label{alg:ga:fit2}
        $ind\_best',\ f\_best' \leftarrow get\_best(pop',\ f\_pop')$ \\ \label{alg:ga:best2}
        \If{$f\_best' < f\_best$}{
            $ind\_best \leftarrow ind\_best'$ \algorithmiccomment{Better individual found} \\
             $f\_best \leftarrow f\_best'$  \\
             $gen\_best\ =\ gen$
        }\label{alg:ga:best2-end}
        $pop, f\_pop \leftarrow environmental\_selection(pop,\ f\_pop,\ pop',\ f\_pop',\ \mu, \ S_e)$ \\ \label{alg:ga:sel}
        $\alpha \leftarrow decay(\alpha,\ \eta,\ gen)$ \\ \label{alg:ga:decay} 
        $gen \leftarrow gen + 1$ \\
    } \label{alg:ga:whilee}

    \textbf{Return} $ind\_best,\ f\_best$ \label{alg:ga:return}
    \caption{Generic GA for network anonymization}
    \label{alg:ga}
\end{algorithm}

\begin{table}[t]
\caption{Hyperparameters used for GA with their symbol (left column), settings used for experiments (center column) and a brief description (right column).}
\label{tab:params}
\begin{tabular}{@{}rcl@{}}
\toprule
Parameter & Setting       & Summary                                                                                          \\ \midrule
$\mu$     & 100           & Initial population size                                                                          \\
$p$       & 0.005            & \begin{tabular}[c]{@{}l@{}}Probability of a bit set to 1\\ in initial population\end{tabular}         \\
$\lambda$ & 150           & Offspring population size                                                                                   \\
$c$       & 25, uniform   & Nr. crossover points                                                                             \\
$\alpha$  & 0.05          & Initial mutation rate                                                                            \\
$\eta$    & 0.001, 0.0025 & Mutation decay rate                                                                              \\
$S_p$        & Roulette wheel            & Parental selection                                                                               \\
$S_e$        & $(\mu + \lambda)$       & Environmental selection                                                                          \\
$\Gamma$  & $0.05 \cdot |E|$     & \begin{tabular}[c]{@{}l@{}}Maximum number of \\ edges to delete\end{tabular}                     \\
$\tau$  & 40            & \begin{tabular}[c]{@{}l@{}}After how many iterations \\ without improvement to stop\end{tabular} \\ \bottomrule
\end{tabular}
\end{table}
\begin{table*}[]
\small
\caption{Table showing network properties (left half) and performance of the algorithm (right half). Properties of social network datasets, detail the number of nodes $|V|$, number of edges $|E|$, average degree, clustering coefficient, fraction of nodes in the largest connected component (LCC), diameter, average distance and initial uniqueness.
Results on reduction in uniqueness for the 4 algorithms (11th to 14th column; higher is better), and improvement factor between the different algorithms (15th to 19th column; higher is better). Result cells marked $\dag$ are not significant. 
}
\label{tab:combined}
\setlength{\tabcolsep}{3.5px}
\begin{tabular}{@{}rrrrrrrrr|rrrrr|rrrrr@{}}
\toprule
                                          & \multicolumn{1}{c}{|V|} & \multicolumn{1}{c}{|E|} & \multicolumn{1}{c}{\begin{tabular}[c]{@{}c@{}}Avg. \\ deg.\end{tabular}} & \multicolumn{1}{c}{\begin{tabular}[c]{@{}c@{}}Clust. \\ coeff\end{tabular}} & \multicolumn{1}{c}{\begin{tabular}[c]{@{}c@{}}Frac \\ LCC\end{tabular}} & \multicolumn{1}{c}{D(G)} & \multicolumn{1}{c}{\begin{tabular}[c]{@{}c@{}}Avg. \\ dist.\end{tabular}} & \multicolumn{1}{c|}{$U(G)$} & \multicolumn{1}{c}{$|V_u|$} & \multicolumn{1}{c}{ES~\cite{romanini2021privacy}} & \multicolumn{1}{c}{UA~\cite{dejong2024anonymization}} & \multicolumn{1}{c}{GA} & \multicolumn{1}{c|}{UGA} & \multicolumn{1}{c}{\begin{tabular}[c]{@{}c@{}}ES \\ vs. \\ GA\end{tabular}} & \multicolumn{1}{c}{\begin{tabular}[c]{@{}c@{}}ES \\ vs.\\ UGA\end{tabular}} & \multicolumn{1}{c}{\begin{tabular}[c]{@{}c@{}}UA \\ vs. \\ GA\end{tabular}} & \multicolumn{1}{c}{\begin{tabular}[c]{@{}c@{}}UA \\ vs. \\ UGA\end{tabular}} & \multicolumn{1}{c}{\begin{tabular}[c]{@{}c@{}}GA \\ vs. \\ UGA\end{tabular}} \\ \midrule
FB Reed98~\cite{networksrepository}       & 962                     & 18,812                  & 39.11                                                                    & 0.33                                                                        & 1.00                                                                    & 6                        & 2.46                                                                      & 0.78                       & 748                          & 34                                                & 38                                                    & 391                    & 391                      & 11.44                                                                       & 11.42                                                                       & 10.43                                                                       & 10.41                                                                        & 1.00 $\dag$                                                                  \\
Blogs \cite{adamic2005political}          & 1,224                   & 16,715                  & 27.31                                                                    & 0.36                                                                        & 1.00                                                                    & 8                        & 2.74                                                                      & 0.49                       & 598                          & 11                                                & 13                                                    & 313                    & 310                      & 29.79                                                                       & 29.50                                                                       & 23.23                                                                       & 23.01                                                                        & 0.9 $\dag$                                                                   \\
FB Simmons~\cite{networksrepository}      & 1,518                   & 32,988                  & 43.46                                                                    & 0.33                                                                        & 1.00                                                                    & 7                        & 2.57                                                                      & 0.79                       & 1,192                        & 26                                                & 41                                                    & 585                    & 567                      & 22.35                                                                       & 21.65                                                                       & 14.15                                                                       & 13.71                                                                        & 0.97 $\dag$                                                                  \\
College msg. \cite{panzarasa2009patterns} & 1,899                   & 13,838                  & 14.57                                                                    & 0.14                                                                        & 1.00                                                                    & 8                        & 3.06                                                                      & 0.24                       & 454                          & 31                                                & 38                                                    & 308                    & 318                      & 10.10                                                                       & 10.43                                                                       & 8.11                                                                        & 8.38                                                                         & 1.03 \hspace{0.142cm}                                                        \\
GRQC collab. \cite{leskovec2007graph}     & 5,242                   & 14,484                  & 5.53                                                                     & 0.69                                                                        & 0.79                                                                    & 17                       & 6.05                                                                      & 0.05                       & 285                          & 2                                                 & 0                                                     & 192                    & 204                      & 100.84                                                                      & 107.16                                                                      & -                                                                           & -                                                                            & 1.06 $\dag$                                                                  \\ \bottomrule
\end{tabular}
\end{table*}

\subsection{Uniqueness-aware Genetic Algorithm (UGA)}\label{sub:aware}
The GA introduced above can, in principle, target all edges in the network for deletion. 
However, deleting certain edges, for example those only affecting nodes that are already anonymous, often does not improve the objective function value.
To be more effective, we instead propose to explicitly target unique edges, focusing on edges connected to at least one unique node as described in Equation~\ref{eq:eu} and illustrated in Figure~\ref{fig:anonymizing}A.
For this, we introduce the uniqueness-aware GA (UGA).

The key difference is in the mutation step in line~\ref{alg:ga:mut} of Algorithm~\ref{alg:ga}.
Based on the set of unique edges $E_u$, as defined  in Equation~\ref{eq:eu}, it assigns a mutation probability $\alpha_{u}(\{v, w\})$ to each edge  $\{v, w\} \in E.$  
This way, the mutation operator only targets edges that are unique and ignores edges connected to anonymous nodes. 

\begin{align}
    \alpha_{u}(\{v, w\}) = 
    \begin{cases}
        \alpha & if\  \{v, w\} \in E_u \\
        0 & otherwise
    \end{cases}
    \label{eq:pua}
\end{align}

This variant is slightly more expensive to compute, as now it is required to compute $E_u$ before the mutation step. 
This computation is essentially as expensive as the objective function itself. 
Then, the probabilities should be assigned to all edges before performing mutation, which has an additional time complexity of $\mathcal{O}(|E|)$. 
\section{Experiments}\label{sec:experiments}
In this section we compare four approaches, being the two GAs introduced in this work, and two baseline measure-agnostic heuristic algorithms from recent literature:
\begin{itemize}
    \item GA: the genetic algorithm proposed in Section~\ref{sub:basic}.
    \item UGA: the uniqueness-aware genetic algorithm proposed in Section~\ref{sub:aware}.
    \item ES: a random edge sampling algorithm, as proposed in \cite{romanini2021privacy}. 
    \item UA: a heuristic algorithm that maximizes the number of affected unique nodes, as proposed in \cite{dejong2024anonymization}. 
\end{itemize}
First, we discuss the experimental setup and real-world network datasets, before extensively comparing the obtained uniqueness of the various algorithms. 
We then assess utility of the anonymized networks, as well as performance on downstream tasks.

\subsection{Experimental setup and hyperparameters}
For the experiments we use the social network datasets in Table~\ref{tab:combined}, which includes a reference for each dataset in the first column.  
These networks vary in topological network properties, such as size, average degree and initial uniqueness.
Given our focus on undirected graphs, timestamps and directionality present in the raw data are ignored in the experiments.

For the GA and UGA, we make use of the parameter settings reported in Table~\ref{tab:params}, resulting in four configurations for which the results on the best configuration are reported.
These parameters are chosen based on extensive hyperparameter tuning experiments on three of the datasets. This process is described extensively in Appendix~\ref{app:tuning}, and is briefly summarized below. 
The hyperparameter tuning process consists of three steps, being 1) successive halving based on 432 initial configurations, 2) running the GA and UGA on the 8 most promising configurations per network, and 3) based on this, choosing four configurations which all networks are run on.

In order to focus on a regime in which data loss is expected to be minimal and data utility can meaningfully be measured, we experiment with a budget $\Gamma$ of 5\% of the edges. 
This choice is based on findings in previous work~\cite{borgatti2006robustness}, which demonstrated that when randomly omitting 5\% of the edges of a social network, topological properties are largely preserved. 

For each of the GAs, we report on the best individual found.
We list values corresponding to the lowest uniqueness for the baseline algorithms. 
Our GAs are implemented in python using NetworkX~\cite{networkx} and available through GitHub\footnote{\url{https://github.com/sambg77/NetworkAnonymization}}.
To compare the GAs to previously proposed approaches from the literature, we use the 
implementation of ES and the UA heuristic used in~\cite{dejong2024anonymization}.
Uniqueness is averaged over 5 runs for each number of edge deletions, and in the process, the uniqueness is recomputed 100 times based on deleted edges in the previous step. Other than that, the default settings are used.

To account for non-determinism of the algorithms used, we average all results over 5 runs, and report on the standard deviation.
To compute utility, we use the igraph and NetworkX libraries~\cite{igraph, networkx}. 
For comparing the similarity of the most central nodes according to betweenness centrality, we report the number of common elements of the top 100 most central nodes in the original network, and the top 100 nodes in the network after anonymization, as described in Section~\ref{sub:utility}.
For community detection, we run the well-known Louvain algorithm~\cite{blondel2008fast} 100 times and then use consensus clustering as implemented in the Netneurotools package~\footnote{\url{https://github.com/netneurolab/netneurotools}}. 

\begin{figure*}[!t]
	\centering
    \includegraphics[width=\textwidth]{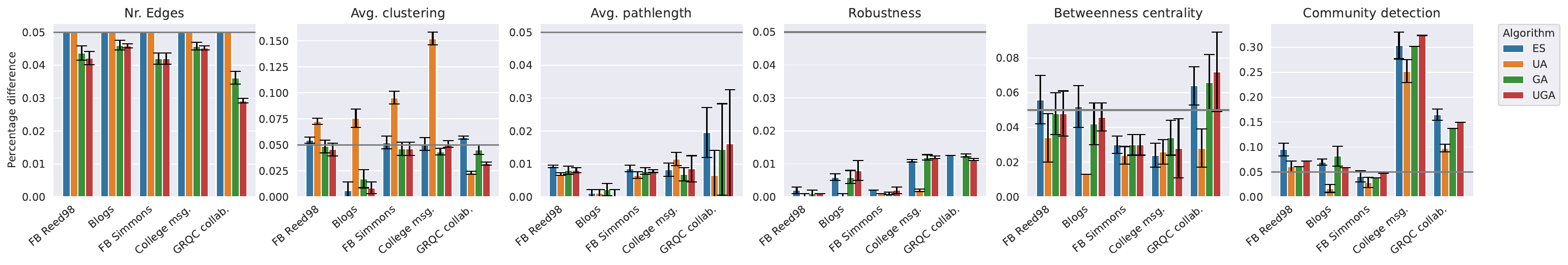}
	\caption{Change in network property (left three figures) or downstream task performance (right three figures) $\pm$ one standard deviation for each network, using the four algorithms (color).}
	\label{fig:util}
\end{figure*}

\subsection{Anonymity}
\label{sub:resano}
The rightmost columns of Table~\ref{tab:combined} show the results of the anonymization experiments consisting of the difference in unique nodes for each algorithm, and a pairwise comparison of the algorithms. 
We performed a Wilcoxon test on all result comparisons, showing that only several results comparing the GA and UGA were not significant; these are marked in Table~\ref{tab:combined}. All other results were significant. 

Overall, the GA always performs significantly better than the baseline heuristics, which do not manage to anonymize many nodes within the 5\% budget.
Whereas the GAs often anonymize hundreds of nodes, the highest amount for the baselines is just 41 nodes.
This was also found in~\cite{dejong2024anonymization}, where it was shown how on some of the networks also included in our analysis, more edge deletions are needed before the uniqueness starts to substantially decrease. 
Results also show how the UA algorithm is not always better than ES; for ``GRQC collab.'' ES manages to anonymize two nodes, while UA anonymizes none within the given budget. 

On average, the GAs anonymize over 14 times more unique nodes compared to the UA baseline. 
Note that to avoid divide-by-zero errors, this excludes the result of ``GRQC collab.'', for which UA failed to anonymize any nodes.
However, in terms of uniqueness reduction, results show similar or identical performance of the GA and the UGA, or, to formulate it differently, the noted differences are mostly not significant. 
One exception is ``College msg.'', for which the UGA actually performs slightly better than the GA. 

\subsection{Utility}
In this section, we compare the two baselines and two GAs on how well they preserve the commonly studied topological network properties described in Section~\ref{sub:utility}. 
The three leftmost results in Figure~\ref{fig:util} show the fraction of edges actually deleted and how the average clustering coefficient and the average shortest path length reduce in value after deleting the selected edges.
In each row, the bold values correspond to the best performing algorithm. 
Overall, most of these best values found are for one of the GAs, showing that their selection method better preserves these topological network properties, although standard deviations are often high, indicating that not all of these results may be considered significant. 
But perhaps more importantly, GAs never show significantly less data utility preservation than baselines, meaning that their superior performance does not come at the price of a reduction in data utility. 

Focusing on the number of edges deleted, we observe that the GAs always delete less than the allowed budget of 5\%. 
For the UGA, on average, fewer edges appear to be deleted than for the GA, although it should be noted that only for ``GRQC collab.'' the difference is larger than the standard deviation.
Moving to the clustering coefficient, we observe that this value changes up to 5.07\% from the original network, which is admissible given that this roughly equals the given budget. 
For the ``GRQC collab.'' network, with a high initial average clustering coefficient of 0.69, UA performs best.
However, for all other networks UA achieves much larger differences compared to the other algorithms, even up to 15.2\%, suggesting that, except in a highly clustered network, the GA performs better at retaining the average clustering coefficient. 
When it comes to the results for average path length, the differences are smaller. 
Often the values are less than 1\% and for some networks, such as ``Blogs'' and ``GRQC collab.'', the standard deviation is quite large, indicating that there is a lot of variation in the solutions.
For ``GRQC collab.'', the values are larger for most algorithms, except for UA, which seems to preserve utility significantly better. 

Overall, the GAs perform better in preserving topological network properties for most networks, as compared to the baselines.
UGA, which requires fewer edge deletions, seems to better preserve network properties on average, although the difference with the generic GA is not always significant. 

\subsection{Downstream tasks}
The three rightmost results of Figure~\ref{fig:util} detail the performance on the downstream tasks discussed in Section~\ref{sub:utility}, comparing the performance on the original network with that on the anonymized networks.

First focusing on the results for robustness, we observe small differences, implying that the giant component stays intact.
The smallest differences are obtained by UA with differences close to zero. 
The GAs perform similarly on these tasks, also similar to ES, which performs worse than UA.
For betweenness centrality, the differences observed are also quite small with at most 0.07, meaning that after anonymization 7 nodes are no longer in the list of the 100 most central nodes.
Again, UA performs best, and the GAs perform similar.
Moving to the last results on community detection, for most networks the NMI is more than 0.9, implying that the communities found before and after anonymization are quite similar.
However, for the ``College msg.'' network, larger differences are observed with NMI values around 0.7.
However, for this network all algorithms fail to preserve the community structure, suggesting that the overall structure is fragile, regardless of what perturbation is applied to it. 

In general, the UA baseline tends to achieve the best performance on downstream tasks, even compared to ES, which deletes edges uniformly at random.
But more important to note from these analyses is that overall, the GA variants often perform similar to the ES baseline, while managing to anonymize on average 14 times more nodes than the best baseline, as we saw in Section~\ref{sub:resano}. 
This suggests that GAs are a good approach if one wishes to maximize anonymization performance, while not affecting data utility more than random approaches would. 


\subsection{Runtime}
\begin{figure}[h]
     \centering
         \includegraphics[width=0.47\textwidth]{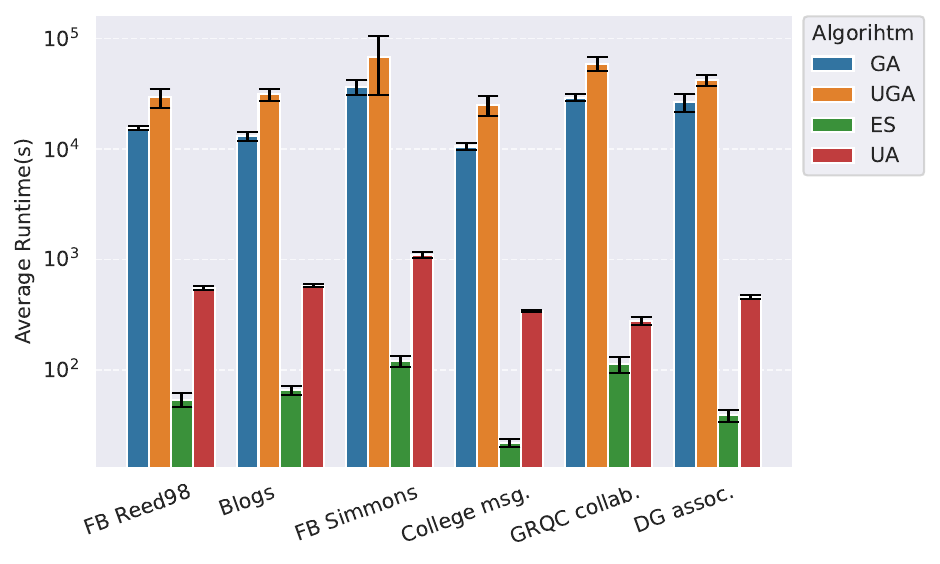}
         \caption{Runtimes for each algorithm (color) on each network dataset (horizontal axis) $\pm$ one standard deviation.}
         \label{fig:runtime}
\end{figure}

Figure~\ref{fig:runtime} shows for each algorithm the runtime $\pm$ one standard deviation on each network. 
First of all, we note that the two GAs require substantially more runtime compared to the much simpler heuristic algorithms.
Comparing the two variants of the GA, we notice that the UGA requires more runtime, due to the additional step that determines which edges are unique.

\section{Conclusions}\label{sec:conc}
In this paper we introduced two genetic algorithms for efficiently anonymizing networks, namely a generic GA and a unique\-ness-aware GA (UGA), which accounts for intricacies specific to the problem of network anonymization.
After extensive hyperparameter optimization, we performed experiments to compare the proposed algorithms against two algorithms from the state-of-the-art. 
We compared the algorithms based on 1) improvement in anonymity, 2) preservation
of network properties, and 3) performance on three downstream tasks.

Results on anonymity indicated how the GAs, on networks with thousands of nodes and tens of thousands of edges, on average anonymized over 14 times more nodes compared to the best baseline algorithm.
In terms of data utility, experimental results showed that GAs perform better or equal compared to the baselines in terms of preserving topological network properties. 
Notably, GAs were better able to preserve the average clustering coefficient. 
Moreover, the GAs always delete less than the allowed budget 5\% of the edges, with the UGA performing slightly better than the GA in this regard. 
On all three downstream tasks commonly performed in social network analysis studied, the performance of the GAs was similar to that of random sampling. 
Overall, our results suggest that GAs are a much better approach if one wishes to maximize anonymization performance while ensuring no substantial damage to data utility.

One direction of future work could be to tailor the GAs to explicitly account for user preferences in terms of retaining performance on downstream tasks. This could be done by incorporating a limit on performance reduction in the objective function by means of penalties. 
To better understand the effect of network measures on GA performance, additional experiments could be done on simulated networks, in which properties such as the clustering coefficient or distance can be artificially controlled. 
Overall, the direction set out in this paper is worth exploring further, as GAs appear a promising method for the network anonymization problem, in particular due to their ability to generate high quality solutions that do not substantially affect data utility.

\begin{acks}
This research was made possible by the Platform Digital Infrastructure SSH (\texttt{\url{http://www.pdi-ssh.nl}}).
We would also like to thank the POPNET team (\texttt{\url{https://www.popnet.io}}) and the Leiden CNS group (\texttt{\url{https://www.computationalnetworkscience.org}}).
\end{acks}

\bibliographystyle{ACM-Reference-Format}
\bibliography{sample-base}

\newpage
\newpage
\clearpage

\appendix
\section{Hyperparameter tuning}\label{app:tuning}
In this appendix, we describe the steps made to find good parameter settings for the network datasets studied in our experiments. 
The set of parameters with which we start can be found in Table~\ref{tab:params_all}.
This results into 432 different configurations.
To find good configurations, we perform the following steps for three of the networks being: ``Blogs'', ``College msg.'' and ``GRQC collab''.
\begin{itemize}
    \item Successive halving (Appendix~\ref{app:halving})
    \item Full run with eight promising configurations (Appendix~\ref{app:second})
    \item Extract four best configurations to use on all networks, and select the best for each network (Appendix~\ref{app:all})
\end{itemize}

\begin{table}[h]
\centering
\small
\caption{Parameter settings used for hyperparameter optimization. }
\label{tab:params_all}
\begin{tabular}{@{}rcl@{}}
\toprule
Symbol    & Settings                 & Parameter                                                                                        \\ \midrule
$\mu$     & 100                      & Initial population size                                                                          \\
$p$       & \{0.25\%, 0.5\%, 2\%\}   & \begin{tabular}[c]{@{}l@{}}Probability bit set to 1\\ in initial population\end{tabular}         \\
$\lambda$ & 150                      & Offspring population size                                                                        \\
$t$       & 5                        & Tournament size                                                                                  \\
$c$       & \{10, 25, 100, uniform\} & Nr. crossover points                                                                             \\
$\alpha$  & \{0.01\%, 0.05\%\}       & Initial mutation rate                                                                            \\
$\eta$    & \{0, 0.001\%, 0.0025\%\} & Mutation decay rate                                                                              \\
$S_p$     & \{RW, TS\}               & Parental selection                                                                               \\
$S_e$     & \{RW, TS, $(\mu + \lambda)$\}      & Environmental selection                                                                          \\
$\Gamma$          & $0.05 \cdot |E|$         & \begin{tabular}[c]{@{}l@{}}Maximum number of \\ edges to delete\end{tabular}                     \\
$\tau$          & 40                       & \begin{tabular}[c]{@{}l@{}}After how many iterations \\ without improvement to stop\end{tabular} \\ \bottomrule
\end{tabular}
\end{table}

\subsection{Successive halving}\label{app:halving}
\begin{figure}[b]
    \centering
    \begin{minipage}{0.35\textwidth}
        \centering
        \includegraphics[width=\textwidth]{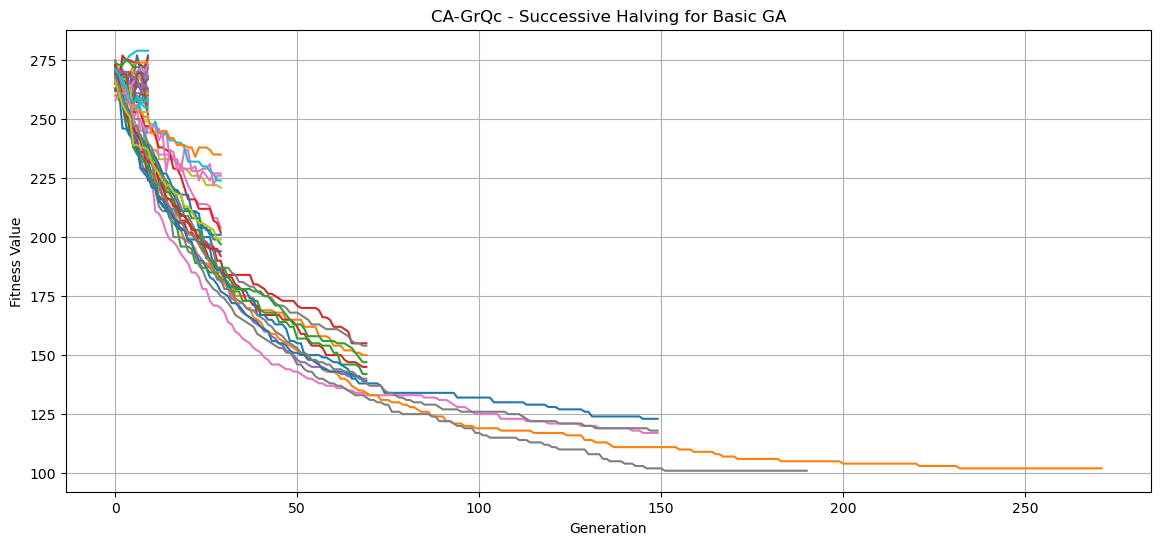}
    \end{minipage}
    \begin{minipage}{0.35\textwidth}
        \centering
        \includegraphics[width=\textwidth]{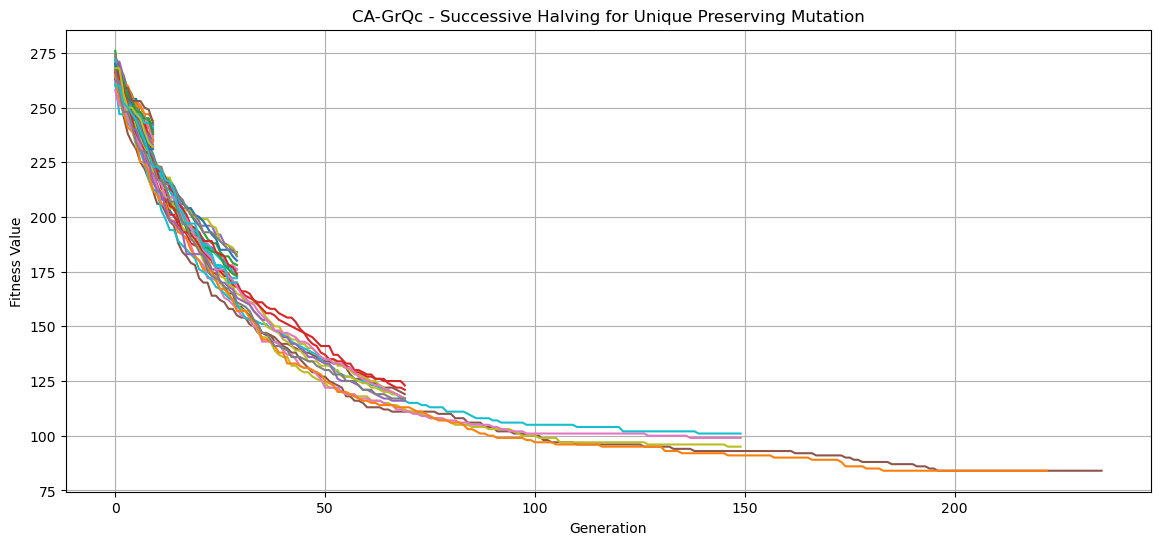}
    \end{minipage}
        \hfill  
    \begin{minipage}{0.35\textwidth}
        \centering
        \includegraphics[width=\textwidth]{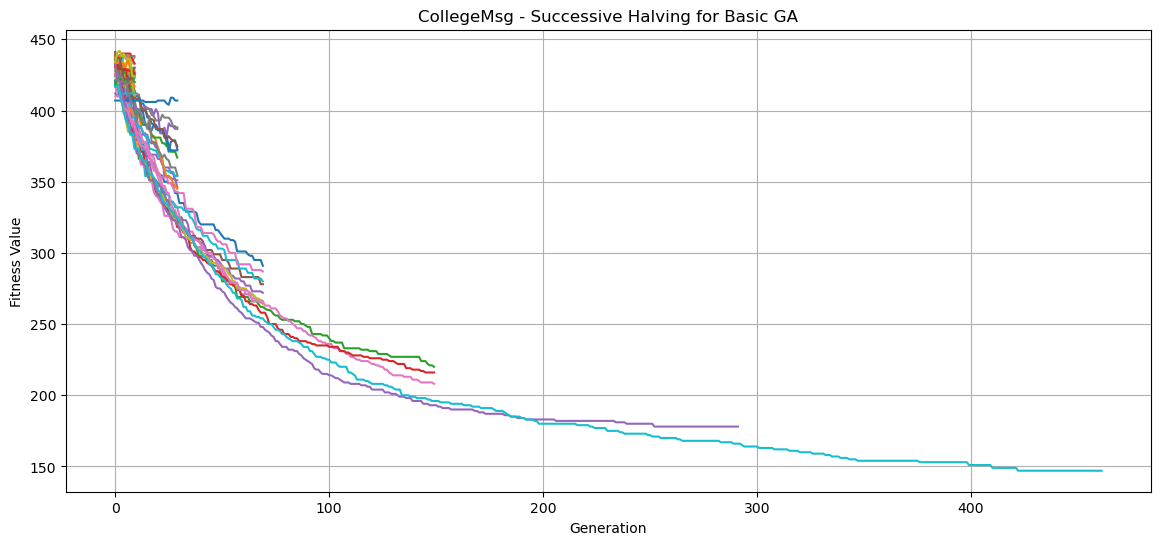}
    \end{minipage}
        \begin{minipage}{0.35\textwidth}
        \centering
        \includegraphics[width=\textwidth]{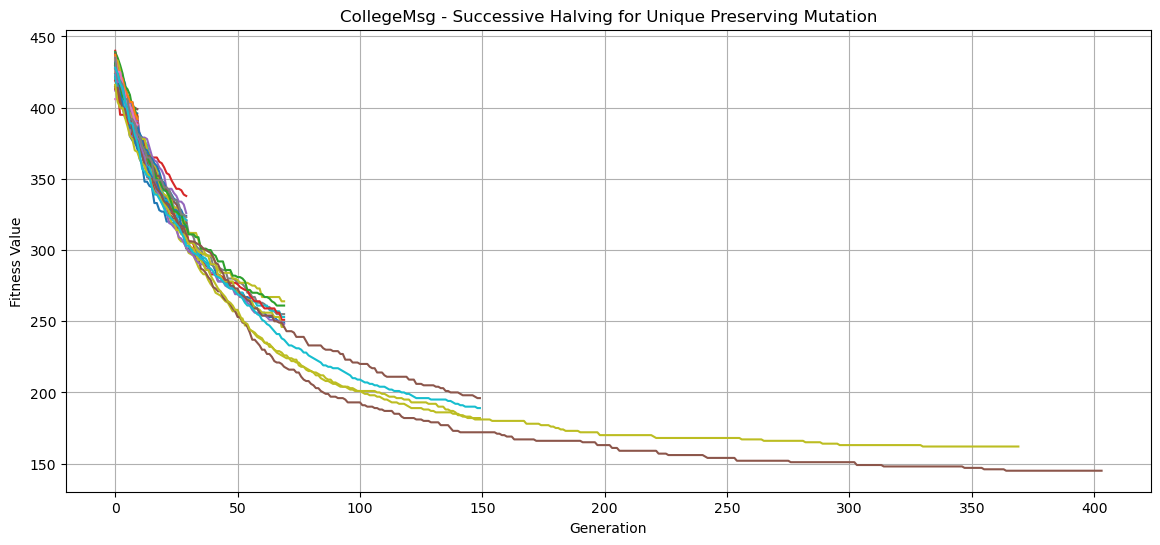}
    \end{minipage}
\hfill

    \begin{minipage}{0.35\textwidth}
        \centering
        \includegraphics[width=\textwidth]{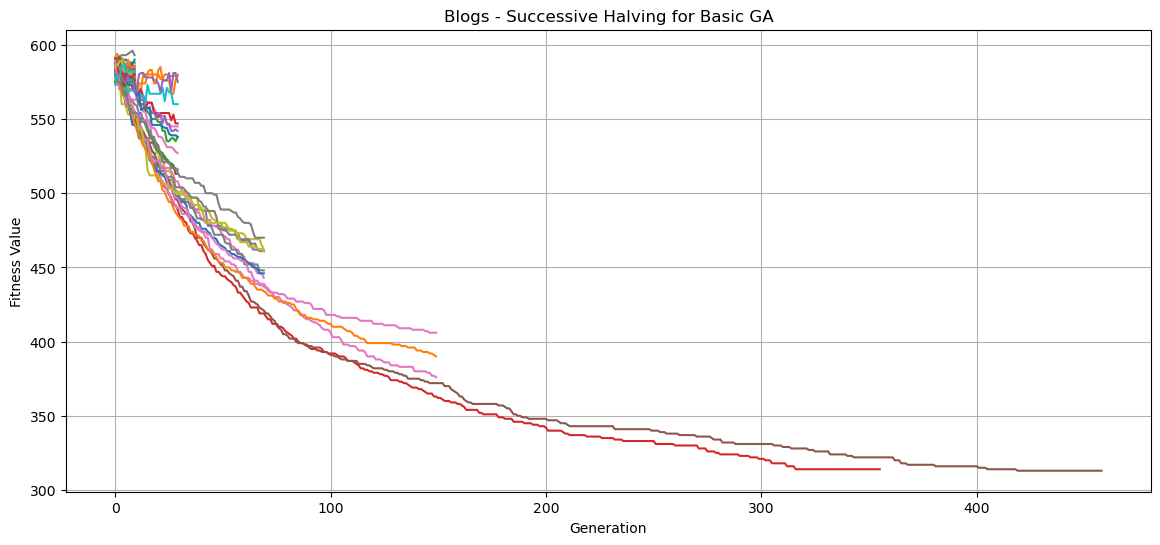}
    \end{minipage}
    \begin{minipage}{0.35\textwidth}
        \centering
        \includegraphics[width=\textwidth]{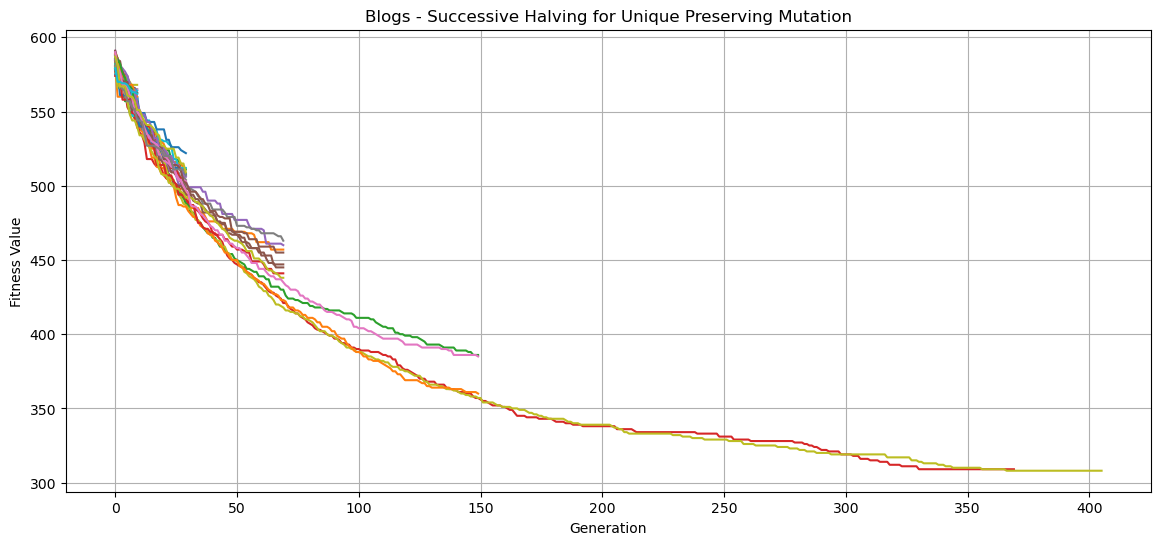}
    \end{minipage}

    \hfill
    \caption{Results obtained for successive halving on the ``GRQC collab.'', ``College mesg.'' and ``Blogs'' networks (rows) using the GA (left column) and UGA (right column). Each line represents a different configuration. The horizontal axis denotes the number of generations, and the vertical axis the number of unique nodes.}
    \label{fig:halving}
\end{figure}

To perform successive halving, we first randomly select 50 configurations. Each 10 generations, we halve the number of configurations run.
Results for the successive halving process for the GA and UGA can be found in Figure~\ref{fig:halving}.

\begin{table*}[h!]
\begin{center}
\footnotesize
\caption{The 3 best configurations for GA and UGA, for each of the three network datasets. Results are averaged over five runs.} 
\label{tab: finalfinal mutation}
\begin{tabular}{cccccccccc}
\hline
\multicolumn{10}{c}{Generic GA (GA)}                                                                                                                                                                                                                                                                                                        \\ \hline
\multicolumn{1}{c|}{}                    & \multicolumn{3}{c|}{GRQC collab.~\cite{leskovec2007graph}}                                                                        & \multicolumn{3}{c|}{College msg.~\cite{panzarasa2009patterns}}                                                                     & \multicolumn{3}{c}{Blogs~\cite{adamic2005political}}                                                       \\ \hline
\multicolumn{1}{c|}{Config.}             & \multicolumn{1}{c|}{3}          & \multicolumn{1}{c|}{4}          & \multicolumn{1}{c|}{6}          & \multicolumn{1}{c|}{4}          & \multicolumn{1}{c|}{6}          & \multicolumn{1}{c|}{7}          & \multicolumn{1}{c|}{4}           & \multicolumn{1}{c|}{5}          & 6           \\ \hline
\multicolumn{1}{c|}{$p$}                 & \multicolumn{1}{c|}{0.25\%}     & \multicolumn{1}{c|}{0.25\%}     & \multicolumn{1}{c|}{0.25\%}     & \multicolumn{1}{c|}{0.5\%}      & \multicolumn{1}{c|}{0.5\%}      & \multicolumn{1}{c|}{0.5\%}      & \multicolumn{1}{c|}{0.5\%}       & \multicolumn{1}{c|}{0.5\%}      & 0.5\%       \\
\multicolumn{1}{c|}{$m$}                 & \multicolumn{1}{c|}{uniform}    & \multicolumn{1}{c|}{25}         & \multicolumn{1}{c|}{25}         & \multicolumn{1}{c|}{10}         & \multicolumn{1}{c|}{uniform}    & \multicolumn{1}{c|}{uniform}    & \multicolumn{1}{c|}{uniform}     & \multicolumn{1}{c|}{uniform}    & 100         \\
\multicolumn{1}{c|}{$\alpha$}            & \multicolumn{1}{c|}{0.05\%}     & \multicolumn{1}{c|}{0.01\%}     & \multicolumn{1}{c|}{0.05\%}     & \multicolumn{1}{c|}{0.01\%}     & \multicolumn{1}{c|}{0.05\%}     & \multicolumn{1}{c|}{0.05\%}     & \multicolumn{1}{c|}{0.05\%}      & \multicolumn{1}{c|}{0.01\%}     & 0.01\%      \\
\multicolumn{1}{c|}{$\eta$}              & \multicolumn{1}{c|}{0.0025\%}   & \multicolumn{1}{c|}{0}          & \multicolumn{1}{c|}{0.0025\%}   & \multicolumn{1}{c|}{0}          & \multicolumn{1}{c|}{0.0025\%}   & \multicolumn{1}{c|}{0.0025\%}   & \multicolumn{1}{c|}{0.0025\%}    & \multicolumn{1}{c|}{0}          & 0           \\
\multicolumn{1}{c|}{$S_p$}           & \multicolumn{1}{c|}{RW}         & \multicolumn{1}{c|}{RW}         & \multicolumn{1}{c|}{RW}         & \multicolumn{1}{c|}{TS}         & \multicolumn{1}{c|}{TS}         & \multicolumn{1}{c|}{RW}         & \multicolumn{1}{c|}{RW}          & \multicolumn{1}{c|}{TS}         & TS          \\
\multicolumn{1}{c|}{$S_e$}           & \multicolumn{1}{c|}{$(\mu + \lambda)$}    & \multicolumn{1}{c|}{$(\mu + \lambda)$}    & \multicolumn{1}{c|}{$(\mu + \lambda)$}    & \multicolumn{1}{c|}{$(\mu + \lambda)$}    & \multicolumn{1}{c|}{$(\mu + \lambda)$}    & \multicolumn{1}{c|}{$(\mu + \lambda)$}    & \multicolumn{1}{c|}{$(\mu + \lambda)$}     & \multicolumn{1}{c|}{$(\mu + \lambda)$}    & $(\mu + \lambda)$     \\ \hline
\multicolumn{1}{c|}{\textbf{$|V_u|$}}     & \multicolumn{1}{c|}{94$\pm$7}   & \multicolumn{1}{c|}{98$\pm$6}   & \multicolumn{1}{c|}{96$\pm$7}   & \multicolumn{1}{c|}{167$\pm$7}  & \multicolumn{1}{c|}{154$\pm$5}  & \multicolumn{1}{c|}{147$\pm$5}  & \multicolumn{1}{c|}{303$\pm$16}  & \multicolumn{1}{c|}{311$\pm$10} & 341$\pm$14  \\ \hline

\multicolumn{10}{c}{Uniqueness-aware GA (UGA)}                                                                                                                                                                                                                                                                                  \\ \hline
\multicolumn{1}{c|}{}                    & \multicolumn{3}{c|}{GRQC collab.~\cite{leskovec2007graph}}                                                                        & \multicolumn{3}{c|}{College msg.~\cite{panzarasa2009patterns}}                                                                     & \multicolumn{3}{c}{Blogs~\cite{adamic2005political}}                                                      \\ \hline
\multicolumn{1}{c|}{Config.}             & \multicolumn{1}{c|}{2}          & \multicolumn{1}{c|}{4}          & \multicolumn{1}{c|}{6}          & \multicolumn{1}{c|}{1}          & \multicolumn{1}{c|}{5}          & \multicolumn{1}{c|}{7}          & \multicolumn{1}{c|}{1}           & \multicolumn{1}{c|}{2}          & 6          \\ \hline
\multicolumn{1}{c|}{$p$}                 & \multicolumn{1}{c|}{0.5\%}      & \multicolumn{1}{c|}{0.5\%}      & \multicolumn{1}{c|}{0.5\%}      & \multicolumn{1}{c|}{2\%}        & \multicolumn{1}{c|}{2\%}        & \multicolumn{1}{c|}{2\%}        & \multicolumn{1}{c|}{2\%}         & \multicolumn{1}{c|}{0.5\%}      & 0.5\%      \\
\multicolumn{1}{c|}{$t$}                 & \multicolumn{1}{c|}{-}          & \multicolumn{1}{c|}{5}          & \multicolumn{1}{c|}{5}          & \multicolumn{1}{c|}{-}          & \multicolumn{1}{c|}{5}          & \multicolumn{1}{c|}{-}          & \multicolumn{1}{c|}{5}           & \multicolumn{1}{c|}{5}          & 5          \\
\multicolumn{1}{c|}{$m$}                 & \multicolumn{1}{c|}{25}         & \multicolumn{1}{c|}{uniform}    & \multicolumn{1}{c|}{uniform}    & \multicolumn{1}{c|}{uniform}    & \multicolumn{1}{c|}{25}         & \multicolumn{1}{c|}{uniform}    & \multicolumn{1}{c|}{uniform}     & \multicolumn{1}{c|}{uniform}    & uniform    \\
\multicolumn{1}{c|}{$\alpha$}            & \multicolumn{1}{c|}{0.05\%}     & \multicolumn{1}{c|}{0.05\%}     & \multicolumn{1}{c|}{0.05\%}     & \multicolumn{1}{c|}{0.01\%}     & \multicolumn{1}{c|}{0.05\%}     & \multicolumn{1}{c|}{0.05\%}     & \multicolumn{1}{c|}{0.05\%}      & \multicolumn{1}{c|}{0.05\%}     & 0.01\%     \\
\multicolumn{1}{c|}{$\eta$}              & \multicolumn{1}{c|}{0.001\%}    & \multicolumn{1}{c|}{0.001\%}    & \multicolumn{1}{c|}{0.0025\%}   & \multicolumn{1}{c|}{0}          & \multicolumn{1}{c|}{0.001\%}    & \multicolumn{1}{c|}{0.001\%}    & \multicolumn{1}{c|}{0.001\%}     & \multicolumn{1}{c|}{0.001\%}    & 0          \\
\multicolumn{1}{c|}{$S_p$}           & \multicolumn{1}{c|}{RW}         & \multicolumn{1}{c|}{TS}         & \multicolumn{1}{c|}{TS}         & \multicolumn{1}{c|}{RW}         & \multicolumn{1}{c|}{TS}         & \multicolumn{1}{c|}{RW}         & \multicolumn{1}{c|}{TS}          & \multicolumn{1}{c|}{TS}         & TS         \\
\multicolumn{1}{c|}{$S_e$}           & \multicolumn{1}{c|}{$(\mu + \lambda)$}    & \multicolumn{1}{c|}{$(\mu + \lambda)$}    & \multicolumn{1}{c|}{$(\mu + \lambda)$}    & \multicolumn{1}{c|}{$(\mu + \lambda)$}    & \multicolumn{1}{c|}{$(\mu + \lambda)$}    & \multicolumn{1}{c|}{$(\mu + \lambda)$}    & \multicolumn{1}{c|}{$(\mu + \lambda)$}     & \multicolumn{1}{c|}{$(\mu + \lambda)$}    & $(\mu + \lambda)$    \\ \hline
\multicolumn{1}{c|}{\textbf{$|V_u|$}}     & \multicolumn{1}{c|}{84$\pm$4}   & \multicolumn{1}{c|}{80$\pm$3}   & \multicolumn{1}{c|}{84$\pm$6}   & \multicolumn{1}{c|}{150$\pm$10} & \multicolumn{1}{c|}{151$\pm$4}  & \multicolumn{1}{c|}{134$\pm$12} & \multicolumn{1}{c|}{293$\pm$12}  & \multicolumn{1}{c|}{298$\pm$5}  & 323$\pm$9  \\ \hline
\end{tabular}
\end{center}
\end{table*}

\subsection{Top three per network}\label{app:second}
Based on the successive halving experiment, we chose a number of configurations. 
However, 
configurations that perform well but take more generations to find good solutions, such as configurations with a mutation decay rate, are likely eliminated.
To account for this, we also included configurations with a mutation decay rate.
We select eight promising configurations for each combination of network and GA variant, for which we run the GAs until termination. 
From these results we select the three best configurations for each combination and report the results in Table~\ref{tab: finalfinal mutation}.

From the obtained results it becomes clear that many parameters within these configurations do not have a large effect on the final uniqueness of the network.
The settings that do seem to affect the results are the number of crossover points $c$ and the mutation decay rate $\eta$.
To extend the experiments to include more networks without elaborate hyperparameter tuning and still obtain good performance, we choose to vary only in these parameters and fix all other parameters.
This results in four configurations reported in Table~\ref{tab:finalconfs}.

\subsection{Final configurations}\label{app:all}
Based on the hyperparameter tuning process, we converged on four configurations. These differ in the number of crossover points $c$ and the mutation decay rate $\eta$ as reported in Table~\ref{tab:finalconfs}.
The results for all networks for the configurations can be found in Table~\ref{tab:top4}.
The best values, indicated in bold, are the results reported in the main paper. 

It is worth noting that the main conclusion of the paper, i.e., that the GAs perform around 14 times better than baselines, do not change substantially between these configurations. 

\begin{table}[b]
\small
\begin{center}
    
\caption{Settings for number of crossover points and mutation decay rate in the final four configurations.}
\label{tab:finalconfs}
\begin{tabular}{@{}rccccl@{}}
\toprule
                    & Conf1    & Conf2    & Conf3   & Conf4  & \\ \midrule
$c$           & Uniform  & 25       & Uniform & 25      & Nr. crossover points \\
$\eta$        & 0.000025 & 0.000025 & 0.00001 & 0.00001 & Mutation decay rate \\ \bottomrule
\end{tabular}
\end{center}

\end{table}

\begin{table}[b]
\begin{center}
\caption{Unique nodes using GA and UGA for the four remaining configurations on all networks. Best values are shown in bold.}
\label{tab:top4}
\setlength{\tabcolsep}{3.5px}
\begin{tabular}{rrlrlrlrl} \hline
GA                                        & \multicolumn{2}{c}{Conf1}   & \multicolumn{2}{c}{Conf2}   & \multicolumn{2}{c}{Conf3}   & \multicolumn{2}{c}{Conf4}   \\ \midrule
FB Reed98~\cite{networksrepository}       & 398          & \textpm 15 & \textbf{357} & \textpm 8  & 386          & \textpm 11 & 359         & \textpm 14  \\
Blogs~\cite{adamic2005political}          & 289          & \textpm 6  & \textbf{285} & \textpm 6  & 308          & \textpm 15 & 312         & \textpm 6   \\
FB Simmons~\cite{networksrepository}      & \textbf{607} & \textpm 15 & 671          & \textpm 19 & 673          & \textpm 17 & 641         & \textpm 29  \\
College msg.~\cite{panzarasa2009patterns} & 153          & \textpm 6  & \textbf{146} & \textpm 3  & 147          & \textpm 3  & 159         & \textpm 7   \\
GRQC collab.~\cite{leskovec2007graph}     & \textbf{93}  & \textpm 6  & 94           & \textpm 11 & 94           & \textpm 5  & 96          & \textpm 3   \\ \midrule
UGA                                       & \multicolumn{2}{c}{Conf1}   & \multicolumn{2}{c}{Conf2}   & \multicolumn{2}{c}{Conf3}   & \multicolumn{2}{c}{Conf4}  \\ \midrule
FB Reed98~\cite{networksrepository}       & 359          & \textpm 5  & \textbf{357} & \textpm 7  & 382          & \textpm 18 & 395         & \textpm 14  \\
Blogs~\cite{adamic2005political}          & \textbf{288} & \textpm 8  & 299          & \textpm 6  & 297          & \textpm 11 & 309         & \textpm 11  \\
FB Simmons~\cite{networksrepository}      & 653          & \textpm 17 & 667          & \textpm 16 & \textbf{625} & \textpm 31 & 686         & \textpm 166 \\
College msg.~\cite{panzarasa2009patterns} & 144          & \textpm 4  & \textbf{136} & \textpm 7  & 145          & \textpm 12 & 146         & \textpm 12  \\
GRQC collab.~\cite{leskovec2007graph}     & 88           & \textpm 6  & 88           & \textpm 3  & 86           & \textpm 3  & \textbf{81} & \textpm 6   \\ \bottomrule
\end{tabular}
\end{center}
\end{table}

\end{document}